\documentstyle[12pt]{article}
\begin{document}
\vskip 0.2cm
\hfill{NIKHEF/97-026}
\vskip 0.2cm
\hfill{ITP-SB-97-35}
\vskip 0.2cm
\hfill{DESY 97-124}
\vskip 0.2cm
\hfill{INLO-PUB-6/97}
\vskip 0.2cm
\centerline{\large\bf {Comparison between the various descriptions for}}
\centerline{\large\bf{charm electroproduction and the HERA-data}}
\vskip 0.2cm
\centerline {\sc M. Buza \footnote{supported by the Foundation
for Fundamental Research on Matter (FOM)}}
\centerline{\it NIKHEF/UVA,}
\centerline{\it POB 41882, NL-1009 DB Amsterdam,}
\centerline{\it The Netherlands}
\vskip 0.2cm
\centerline {\sc Y. Matiounine }
\centerline{\it Institute for Theoretical Physics,}
\centerline{\it State University of New York at Stony Brook,}
\centerline{\it New York 11794-3840, USA}
\vskip 0.2cm
\centerline {\sc  J. Smith \footnote{on leave from ITP, SUNY at Stony Brook,
New York 11794-3840, USA} }
\centerline{\it Deutsches Electronen-Synchrotron DESY}
\centerline{\it Notkestrasse 85, D-22603, Hamburg}
\centerline{\it Germany}
\vskip 0.2cm
\centerline {\sc W.L. van Neerven}
\centerline{\it Instituut-Lorentz,}
\centerline{\it University of Leiden,}
\centerline{\it PO Box 9506, 2300 RA Leiden,}
\centerline{\it The Netherlands}
\vskip 0.2cm
\centerline{July 1997}
\vskip 0.2cm
\centerline{\bf Abstract}
\vskip 0.3cm
We examine the charm component $F_{2,c}(x,Q^2,m^2)$
of the proton structure function $F_2(x,Q^2)$ in three 
different schemes and compare
the results with the data in the $x$ and $Q^2$ region explored by the
HERA experiments. Studied are (1) the three flavour number scheme (TFNS) 
where the production mechanisms are given by the photon-gluon fusion process  
and the higher order reactions with 
three light-flavour parton densities as input 
(2) the four flavour number scheme (FFNS) where $F_{2,c}$ is expressed in  
four light flavour densities including one for the charm quark
and (3) a variable-flavour number scheme (VFNS) which interpolates between the 
latter two. Both the VFNS and the TFNS give good descriptions of the
experimental data. However one cannot use the FFNS for the 
description of the data at small $Q^2$.

\vfill
\newpage

The study of charm production in deep-inelastic electron-proton scattering
has become an important issue in the extraction of parton densities
in the proton. The reason is that the charm content  
$F_{2,c}(x,Q^2,m^2)$ ($m$ is the mass of the charm quark)
in the proton structure function $F_2(x,Q^2)$ has grown from
around one percent in the $x$ and $Q^2$ region covered by the 
EMC experiment \cite{emc} to around twenty-five percent in the
$x$ and $Q^2$ region covered by the H1 \cite{h1} and ZEUS
\cite{zeus} experiments at HERA. Therefore the analysis of data
to yield parton densities can no longer treat charm electroproduction
as a small correction.

Let us begin with a brief review of the QCD contributions to 
$F_{2,c}(x,Q^2,m^2)$. 
In lowest order (LO) the charm quark pair is produced via the photon-gluon
fusion mechanism which implies that the Born contribution only contains a
gluon in the initial state (extrinsic production). 
In next-to-leading order (NLO) we have in addition to the
above reaction also Bethe-Heitler and Compton subprocesses containing 
the three light flavours u, d, s and their antiparticles
${\bar {\rm u}}$, ${\bar{\rm d}}$, ${\bar {\rm s}}$ in the initial state.
We will call this the three flavour number scheme (TFNS) description 
for charm production. The NLO calculations \cite{lrsn} reveal 
that the photon-gluon process dominates the 
other production mechanisms so that charm electroproduction 
yields a measurement of the gluon density, in particular at small $x$.
Generalizing the TFNS to all orders in $\alpha_s$
we obtain the following formula for the charm
content of the proton structure function
\begin{eqnarray}
F_{2,c}^{\rm EXACT}(3,x,Q^2,m^2)& =&
x \int_x^{z_{\rm max}}\frac{dz}{z}\Biggl\{
\frac{1}{3} \sum\limits_{k=1}^{3} e_k^2 
\Biggl[
\Sigma\Big(3,\frac{x}{z},\mu^2\Big)  
L_{2,q}^{\rm S}\Big(3,z,\frac{Q^2}{m^2},\frac{m^2}{\mu^2}\Big)
\nonumber\\
&& 
+ G\Big(3,\frac{x}{z},\mu^2\Big) 
L_{2,g}^{\rm S}\Big(3,z,\frac{Q^2}{m^2},\frac{m^2}{\mu^2}\Big)
\nonumber\\
&& + 3 \Delta\Big(3,\frac{x}{z},\mu^2\Big)
L_{2,q}^{\rm NS}\Big(3,z,\frac{Q^2}{m^2}, \frac{m^2}{\mu^2}\Big) \Biggr]
\Biggr\}
\nonumber\\
&&+ x e_c^2 
\int_x^{z_{\rm max}} \frac{dz}{z}
\Biggl[ \Sigma\Big(3,\frac{x}{z},\mu^2\Big)  
H_{2,q}^{\rm S}\Big(3,z,\frac{Q^2}{m^2},\frac{m^2}{\mu^2}\Big)
\nonumber \\ &&
+  G\Big(3,\frac{x}{z},\mu^2\Big)  
H_{2,g}^{\rm S}\Big(3,z,\frac{Q^2}{m^2},\frac{m^2}{\mu^2}\Big)
\Biggr] \,,
\end{eqnarray}
where $e_c$ stands for the charge of the charm quark and the number of light
flavours is three ($n_f =3$). The summation variable $k$ refers 
to the light quarks u,d and s.
The variable $z$ is the partonic longitudinal momentum fraction
which has a maximum value $z_{\rm max} = Q^2/(Q^2 + 4 m^2)$.
The function $\Delta$ is the non-singlet (NS) 
(with respect to the flavour group)
combination of light flavour densities, which are functions
of the scale $\mu$. The function $\Sigma$ is the singlet (S)
combination of these densities
while $G$ stands for the density of the gluon. Further
$L_{2,k}$ and $H_{2,k} ( k=q,g )$ represent the charm-quark coefficient
functions which can also be separated into flavour singlet and  
flavour non-singlet parts. The functions $L_{2,k}$
describe the reactions where the virtual photon couples to the light
quarks (u, d, s, ${\bar {\rm u}}$, ${\bar {\rm d}}$ and ${\bar {\rm s}}$),
whereas the $H_{2,k}$ describe the reactions 
where the virtual photon couples to the c ($\bar {\rm c}$) quark. 
Hence $L_{2,k}$ and $H_{2,k}$ in Eq. (1) are multiplied by $e_k^2$ 
and $e_c^2$ respectively. 
Moreover, when the reaction where the photon couples
to the charm quark contains a light quark in the initial 
state, then it can only proceed via the exchange of a gluon 
in the $t$-channel. Therefore $H_{2,q}$ is a flavour singlet.
This is in contrast with $L_{2,q}$ 
which has both flavour singlet and non-singlet contributions. 
Finally we have to add to Eq. (1) that part of the proton structure function 
$F_2(x,Q^2)$ which contains charm quark
loop contributions to the gluon self energies. The latter are only
inserted in the matrix elements containing light partons (u,d,s,g). 
Combining this part of $F_2(x,Q^2)$
with the contributions from $L_{2,k}$ leads to the correct 
asymptotic behaviour of the charm quark coefficient functions
at $Q^2 \gg m^2$ which allows us to perform the mass factorization 
discussed below.
The NLO contributions to the charm quark coefficient functions 
were originally calculated in \cite{lrsn}. 
These functions were available analytically
in the form of two-dimensional integrals which were then computed numerically.
To speed up the integrations in Eq. (1) a two-dimensional
grid of values for $L_{2,k}$ and $H_{2,k}$ together with an interpolation 
routine was provided in \cite{rsn}. 

As a check on our exact formulae we
have also calculated the asymptotic expressions
for $L_{2,k}$ and $H_{2,k}$ in the limit $Q^2 \gg m^2$
\cite{bmsmn}. These asymptotic charm quark coefficient
functions contain the large logarithms of the type 
$\ln^i(Q^2/m^2)\ln^j(\mu^2/m^2)$
which dominate charm production far away from threshold. This has been 
shown in \cite{bmsn} by numerically 
comparing the asymptotic structure function defined
by
\begin{equation}
F_{2,c}^{\rm ASYMP}(3,x,Q^2, m^2 )= \lim_{Q^2 \gg m^2}
\Big[F_{2,c}^{\rm EXACT}(3,x,Q^2,m^2)\Big] \,,
\end{equation}
with the exact structure function in Eq. (1). Notice
that in the last equation the exact charm quark coefficient functions 
are replaced by their asymptotic
analogues mentioned above and $z_{\rm max}=1$. 
The NLO results \cite{bmsn} reveal
that for $Q^2 > 20~({\rm GeV/c})^2$ 
and $x<0.01$, $F_{2,c}^{\rm ASYMP}$ coincides
with $F_{2,c}^{\rm EXACT}$ which implies that the large logarithms 
mentioned above entirely determine the charm component of the 
structure function. 
Since these corrections vitiate the
perturbation series when $Q^2$ gets large they should be resummed in all 
orders of perturbation theory. This procedure has been carried out in
\cite{bmsn} and it consists of four steps. First we add 
the light parton (u,d,s,g) component of the proton
structure function defined by $F_2(3,x,Q^2)$
to $F_{2,c}^{\rm ASYMP}$ in Eq. (2). 
Second we apply mass 
factorization to the asymptotic charm quark coefficient functions containing
the logarithmic terms $\ln^i(Q^2/m^2)\ln^j(\mu^2/m^2)$. In this way 
the $m^2$-dependence is separated from the $Q^2$-dependence. 
The logarithms in the
charm quark mass are put in the charm quark operator matrix elements 
whereas the logarithms in the variable $Q^2$ are
transferred to the light parton coefficient functions denoted by 
${\cal C}_{2,k}$ ($k=q,g$). Both quantities have been calculated up to order 
$\alpha_s^2$ in \cite{bmsmn}, \cite{bmsn} and \cite{zn1} respectively.
In the third step we define new parton densities in a four flavour number
scheme (FFNS) which can be written as convolutions of the original parton
densities in the three flavour number scheme (TFNS) 
with the charm quark operator matrix elements mentioned above.
Therefore we obtain a new parton density which represents the charm quark
and is denoted by $f_{c + \bar c}(z,\mu^2)$. 
The latter has the property that it
does not vanish at $\mu = m$ in the $\overline{\rm MS}$-scheme contrary to
what is usually assumed in the literature. In the fourth step we rearrange
terms and obtain
$F_{2,c}^{\rm ASYMP}(3,x,Q^2,m^2)+F_2(3,x,Q^2)=F_2(4,x,Q^2)$
which is the FFNS result for the proton structure function. From the latter 
quantity one can extract the expression for the charm quark component of
the proton structure function in the FFNS which will be denoted by
\begin{eqnarray}
F_{2,c}^{\rm PDF} (4,x, Q^2) &  = &  e_c^2 \int_x^{1} \frac{dz}{z}
\Big[f_{c +\bar c}\Big(4,\frac{x}{z},\mu^2\Big) 
{\cal C}_{2,q}^{\rm NS}\Big(4,z,\frac{Q^2}{\mu^2}\Big)
\nonumber\\
&& + \Sigma\Big(4,\frac{x}{z},\mu^2\Big) \tilde{\cal C}_{2,q}^{\rm PS}
\Big(4,z,\frac{Q^2}{\mu^2}\Big)
\nonumber \\ &&
 + G\Big(4,\frac{x}{z},\mu^2\Big) \tilde{\cal C}_{2,g}^{\rm S}
\Big(4,z,\frac{Q^2}{\mu^2}\Big) \Big]\,,
\end{eqnarray}
where we have defined
\begin{eqnarray}
{\cal C}_{2,q}^{\rm S}(n_f)={\cal C}_{2,q}^{\rm NS}(n_f)
+n_f \tilde{\cal C}_{2,q}^{\rm PS}(n_f) \,, \qquad
{\cal C}_{2,g}^{\rm S}=n_f\tilde{\cal C}_{2,g}^{\rm S} \,.
\end{eqnarray} 
The superscript PDF in Eq. (3) stands for parton density function which means
that the charm component of the structure function can be completely
expressed into parton densities multiplied by the light parton coefficient
functions. Notice that $F_{2,c}^{\rm PDF}$ is a renormalization group
invariant which implies that this structure function, like the one in Eq. (1)
is explicitly independent of $\mu$ so that it becomes a physical quantity. 
Further it originates from the charm quark coefficient functions $H_{2,k}$
since both are proportional to $e_c^2$. The functions $L_{2,k}$, which are
proportional to the light charge squared $e_k^2$, also contribute to 
$F_2(4,x,Q^2)$ where they increase the number of flavours by one,
and provide us with order $\alpha_s^2$ matching relations 
between the TFNS and FFNS light parton densities.
The FFNS charm quark density is mainly determined by
the size of the TFNS gluon density $G(3,z,\mu^2)$. 
Therefore the latter plays a major role in
the behaviour of $F_{2,c}^{\rm EXACT}$ in Eq. (1) 
as well as $F_{2,c}^{\rm PDF}$ in Eq. (3). 
An analysis of both structure functions in \cite{bmsn} revealed
that the former gives the best description of charm production in the
threshold region where $Q^2$ is small and $x$ is large. On the other hand
when $Q^2$ is large and $x$ is small it turns out that it is better to use
$F_{2,c}^{\rm PDF}$ because it is in this region where the large logarithms
$\ln^i(Q^2/m^2)\ln^j(\mu^2/m^2)$ dominate so that they have to be resummed.
Therefore the TFNS is the most suitable scheme for the charm
component of the structure function near threshold whereas far away from this
region it turns out that the FFNS is more appropriate. 

One also needs a scheme which merges the advantages of these 
two pictures and provides us with good description of $F_{2,c}$ 
in the intermediate regime in $Q^2$. This is
given by the so called variable flavour number scheme (VFNS). 
In \cite{bmsn} we proposed the following VFNS structure function
\begin{eqnarray}
F_{2,c}^{\rm VFNS}(x,Q^2,m^2) &=&
F_{2,c}^{\rm PDF}(4,x,Q^2)
+ F_{2,c}^{\rm EXACT}(3,x,Q^2,m^2) \nonumber \\ &&
- F_{2,c}^{\rm ASYMP}(3,x,Q^2,m^2)\,.
\end{eqnarray}
The above expression is a generalization of Eq. (9) in \cite{acot}, which
was only presented in LO and has been implemented in a recent
global parton density analysis \cite{lt}. 
(A different VFNS scheme has recently been proposed in
\cite{mrrs}.) In \cite{acot} and in \cite{or} it was shown 
that $F_{2,c}^{\rm VFNS}$
in LO is less sensitive to variations in the scale $\mu$ than each
term on the right-hand-side of Eq. (5) separately.  
Further the VFNS scheme in LO has the properties
that for $Q^2 \gg m^2$, $F_{2,c}^{\rm EXACT} \rightarrow F_{2,c}^{\rm ASYMP}$
which means that $F_{2,c}^{\rm VFNS} \rightarrow F_{2,c}^{\rm PDF}$
while at low
$Q^2$ (i.e. $Q^2 \leq m^2$) $F_{2,c}^{\rm ASYMP} \rightarrow 
F_{2,c}^{\rm PDF}$ 
\footnote{This is only true if we put $z_{\rm max}=1$
for $F_{2,c}^{\rm ASYMP}$, which was not done in \cite{bmsn}.}
so that $F_{2,c}^{\rm VFNS} \rightarrow F_{2,c}^{\rm EXACT}$.
However the last relation is no longer true in higher order in $\alpha_s$.
The first reason is that all parton densities and the running
coupling constant in $F_{2,c}^{\rm PDF}$ 
have four flavours whereas those appearing in
$F_{2,c}^{\rm EXACT}$ and $F_{2,c}^{\rm ASYMP}$ have only three flavours. 
Notice that in lowest order we are allowed to 
put $\alpha_s(3,\mu^2)=\alpha_s(4,\mu^2)$
and $G(3,z,\mu^2)=G(4,z,\mu^2)$. Hence in LO we can use the
same number of flavours in all structure functions (PDF, EXACT and ASYMP).
In NLO this is no longer true, which
implies that there are regions in (small) $Q^2$ where 
$F_{2,c}^{\rm ASYMP}$ cannot exactly cancel $F_{2,c}^{\rm PDF}$.
The second reason is that new production mechanisms appear in NLO leading 
to the coefficient functions $L_{2,k}$. The latter contain the 
prefactor $e_k^2$ (e.g. the Compton process) and show up
in $F_{2,c}^{\rm EXACT}$ and $F_{2,c}^{\rm ASYMP}$ but not in
$F_{2,c}^{\rm PDF}$, which is proportional to $e_c^2$ only (see Eq. (3)).
Fortunately it turns out that both these corrections  
are small for $Q^2 \leq m^2$ so 
that $F_{2,c}^{\rm VFNS,(2)} \approx F_{2,c}^{\rm EXACT,(2)}$
numerically. 
Yet another problem arises when one chooses a parton density set in which the
charm quark density vanishes at $\mu=m$. As we have found in \cite{bmsn}
this property does not hold anymore beyond order $\alpha_s$ in the
$\overline{\rm MS}$-scheme. This can be remedied by either making a new
parton density set satisfying the conditions presented in 
Eqs. (2.37)-(2.41) in \cite{bmsn} or by choosing a different
scheme. As we use charm quark densities
satisfying the condition $f_{c + \bar c}(4,z,m^2)=0$ then we have to make an
oversubtraction, which implies that the light flavour coefficient functions 
$\tilde {\cal C}_{2,g}^{\rm S}$ and $\tilde {\cal C}_{2,q}^{\rm PS}$ in
Eq. (3) differ from the results obtained in the $\overline{\rm MS}$-scheme.

As an application we have studied the charm component of the proton structure
function in the three schemes mentioned above. The coefficient functions used
for these structure functions are all computed up to order $\alpha_s^2$ 
(see \cite{lrsn}, \cite{bmsmn}, \cite{zn1}) so that we will denote them by
$F_{2,c}^{\rm EXACT,(2)}$ in Eq. (1), $F_{2,c}^{\rm PDF,(2)}$ in Eq. (3) 
and $F_{2,c}^{\rm VFNS,(2)}$ in Eq. (5). In Figs. 1-4 we have made plots
of the $x$ dependence of these functions for 
$Q^2= 3,12,45$ and 170 $({\rm GeV/c})^2$ respectively and compared 
the results with the recent data
from the H1 \cite{h1} and ZEUS \cite{zeus} collaborations.
We have chosen the next-to-leading log 
FFNS parametrization in \cite{grv92} 
with the condition that $f_{c + \bar c}(z,\mu^2)=0$ for $\mu \leq m$
for the input parton density set. 
Further we have taken the two-loop corrected running coupling constant with
$\Lambda_4= 200$ MeV. Notice that in principle one has to choose a TFNS
parametrization for both the parton densities (see e.g
\cite{grv94}) and the running coupling constant for the computation of
$F_{2,c}^{\rm EXACT}$ and $F_{2,c}^{\rm ASYMP}$. However we have checked that
the latter are not significantly altered when 
we replace the parton densities in \cite{grv94}
by those in \cite{grv92}.  
Finally we adopt the mass factorization scale 
from \cite{acot} and use
\begin{eqnarray}
\mu^2 \!\!\!\!\!\!&& 
=\,\,\,m^2+ k Q^2 (1- m^2/Q^2)^n \quad \mbox{for} \quad Q^2 > m^2\,,
\nonumber\\
&& =\,\,\, m^2 \quad \mbox{for} \quad Q^2 \leq m^2\,,
\end{eqnarray}
with $k=0.5$, $n=2$ and $m = 1.5~({\rm GeV}/c^2)$.\\
At $Q^2= 3$ $({\rm GeV/c})^2$ (Fig. 1) the structure function 
$F_{2,c}^{\rm PDF,(2)}$
is negative over the whole $x$-region so that we have not shown it.
Nevertheless it is almost exactly equal to $F_{2,c}^{\rm ASYMP,(2)}$ 
so these terms cancel in Eq. (5). Therefore at low $Q^2$ the last two structure
functions have no physical meaning because the parts in the heavy quark
coefficient functions leading to the large logarithms do not dominate the
QCD corrections to the structure function in Eq. (1).
Hence at low $Q^2$ the 
correct description is given
by $F_{2,c}^{\rm EXACT,(2)}$ (TFNS)  which is almost equal to 
$F_{2,c}^{\rm VFNS,(2)}$ so that it is not necessary to use the VFNS 
description. However for larger $Q^2$, i.e.
$Q^2 \geq 12$ $({\rm GeV/c})^2$, both $F_{2,c}^{\rm EXACT,(2)}$ and
$F_{2,c}^{\rm PDF,(2)}$ are in agreement with most of the data.
The same holds for $F_{2,c}^{\rm VFNS,(2)}$ which lies between the previous
two structure functions.
The only exception is the value $Q^2 = 170$ $({\rm GeV}/c^2)$
(see Fig. 4), where all descriptions are poor.
The plots show the general features
that $F_{2,c}^{\rm EXACT,(2)}$ (TFNS) is always below $F_{2,c}^{\rm PDF,(2)}$
(FFNS) which is due to the resummation of all the large 
logarithms in the latter
structure function. Notice that these large terms are only included up to
finite order in $F_{2,c}^{\rm EXACT,(2)}$ and $F_{2,c}^{\rm ASYMP,(2)}$.
Finally as expected 
$F_{2,c}^{\rm VFNS,(2)} \approx F_{2,c}^{\rm EXACT,(2)}$ 
at large $x$ (threshold region)
while $F_{2,c}^{\rm VFNS,(2)} \approx F_{2,c}^{\rm PDF,(2)}$ 
at small $x$. However it is clear that one needs
more precise data in finer bins of $x$ and large $Q^2$ to discriminate
between the TFNS and the FFNS in order to observe the resummation effects
incorporated into $F_{2,c}^{\rm PDF,(2)}$. The same holds for medium $Q^2$
where one would like to distinguish between the VFNS and the other two
schemes.\\[5mm]
\noindent
Acknowledgements\\

The research of J. Smith and Y. Matiounine was partially 
supported by the contract NSF 93-09888. J. Smith would like to thank the
Alexander von Humbolt Stiftung for an award to allow him to spend his
Sabbatical leave in DESY.

%

\centerline{\bf \large{Figure Captions}}
\begin{description}
\item[Fig. 1.]
$F^{{\rm VFNS},(2)}_{2,c}(x,Q^2)$ solid line
and $F^{{\rm EXACT},(2)}_{2,c}(x,Q^2)$ dotted line
plotted as functions of $x$ at $Q^2 =3$ $({\rm GeV}/{\rm c})^2$.
The experimental point is from \cite{zeus}.
\item[Fig. 2.]
$F^{{\rm VFNS},(2)}_{2,c}(x,Q^2)$ solid line,
$F^{{\rm PDF},(2)}_{2,c}(x,Q^2)$ dashed line and
$F^{{\rm EXACT},(2)}_{2,c}(x,Q^2)$ dotted line 
plotted as functions of $x$ at $Q^2 =12$ $({\rm GeV}/{\rm c})^2$. 
The experimental points are from \cite{h1} closed circles 
and \cite{zeus} open circles.
\item[Fig. 3.]
Same as Fig. 2 for $Q^2 = 45$ $({\rm GeV}/{\rm c})^2$.

\item[Fig. 4.]
Same as Fig. 2 for $Q^2 = 170$ $({\rm GeV}/{\rm c})^2$.
The experimental point is from \cite{zeus}.

\end{description}
\setlength{\unitlength}{1cm}
\begin{figure}[p]
\begin{center}
\begin{picture}(15,20)
\put(-2,0){\includegraphics{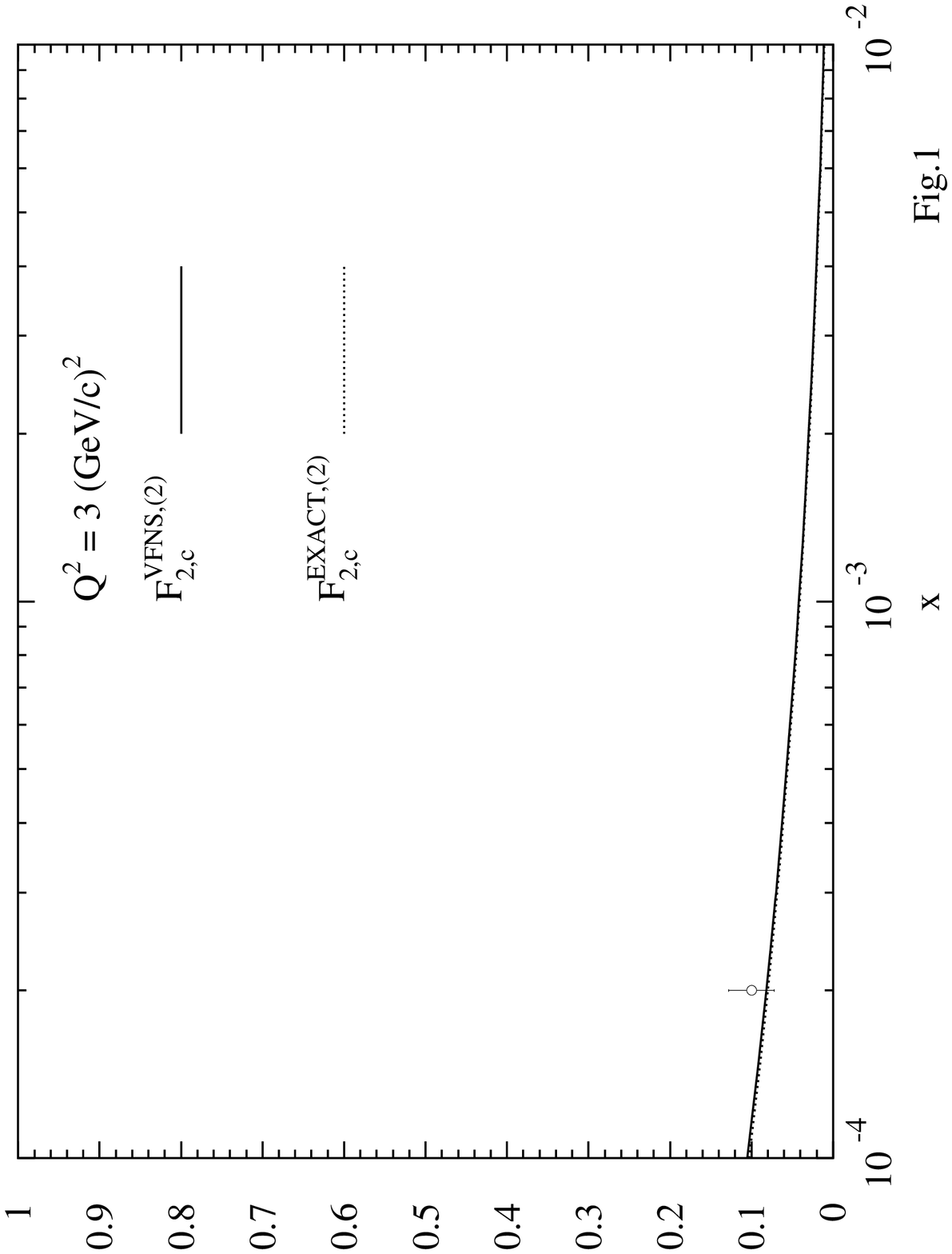}}
\end{picture}
\end{center}
\end{figure}
\begin{figure}[p]
\begin{center}
\begin{picture}(15,20)
\put(-2,0){\includegraphics{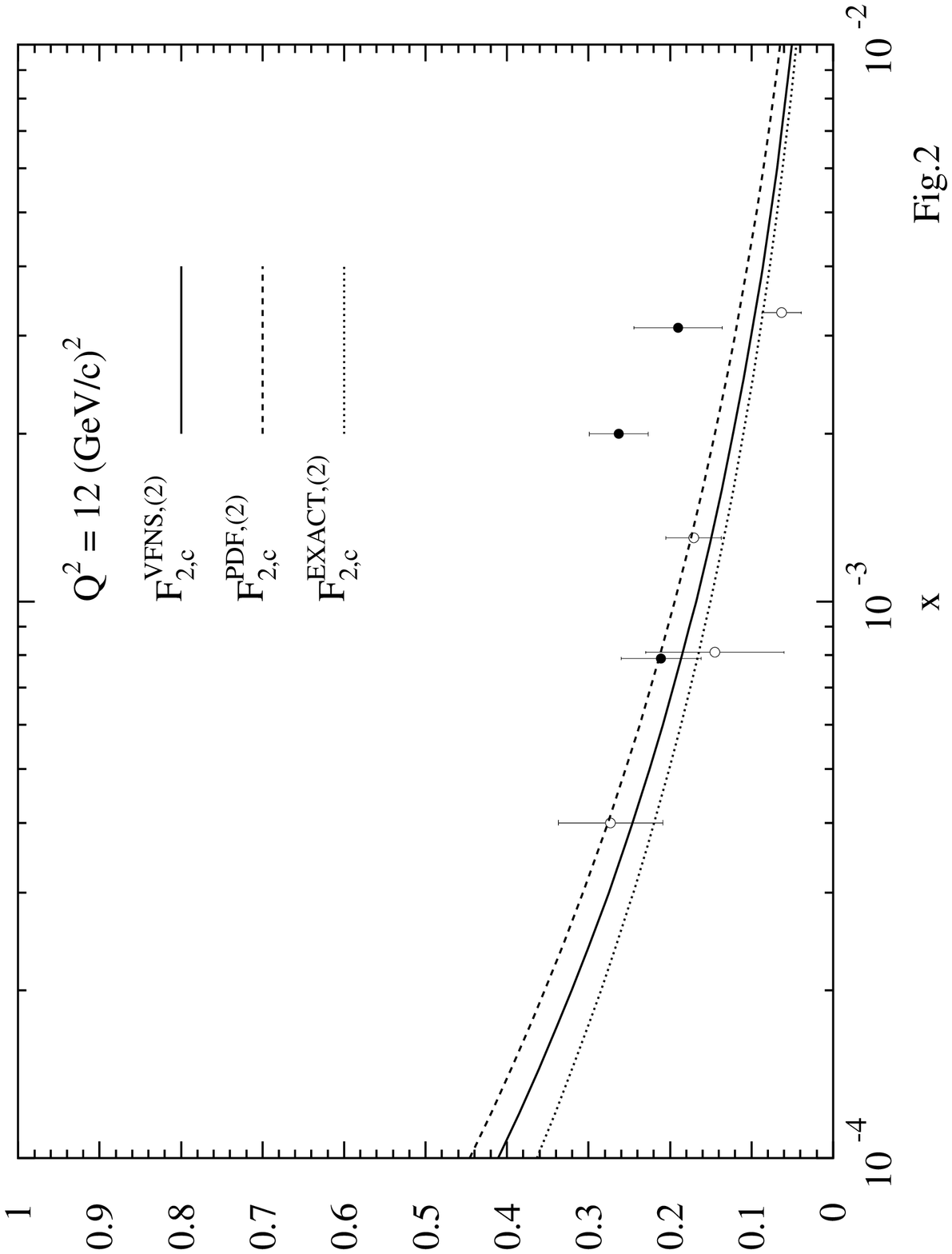}}
\end{picture}
\end{center}
\end{figure}
\begin{figure}[p]
\begin{center}
\begin{picture}(15,20)
\put(-2,0){\includegraphics{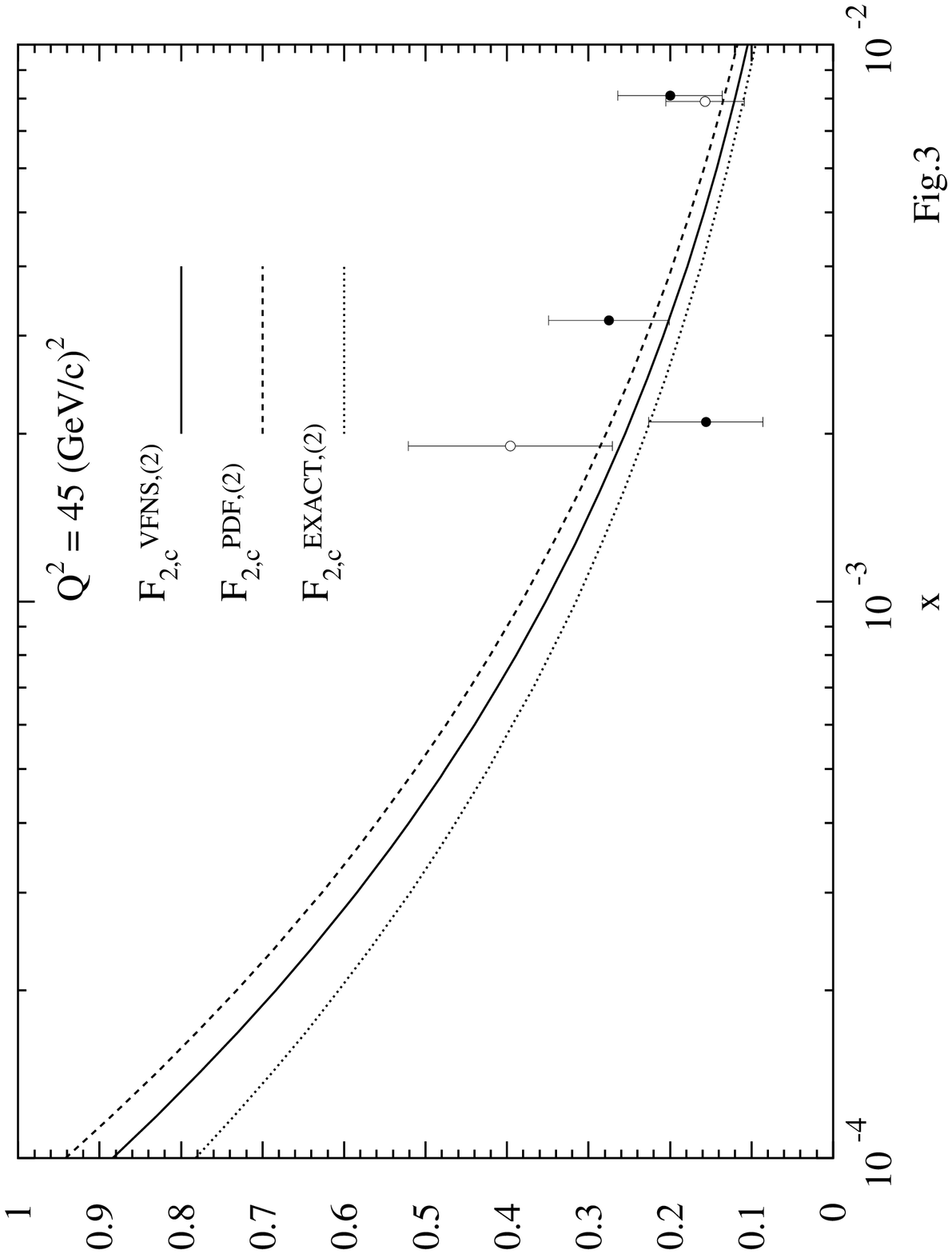}}
\end{picture}
\end{center}
\end{figure}
\begin{figure}[p]
\begin{center}
\begin{picture}(15,20)
\put(-2,0){\includegraphics{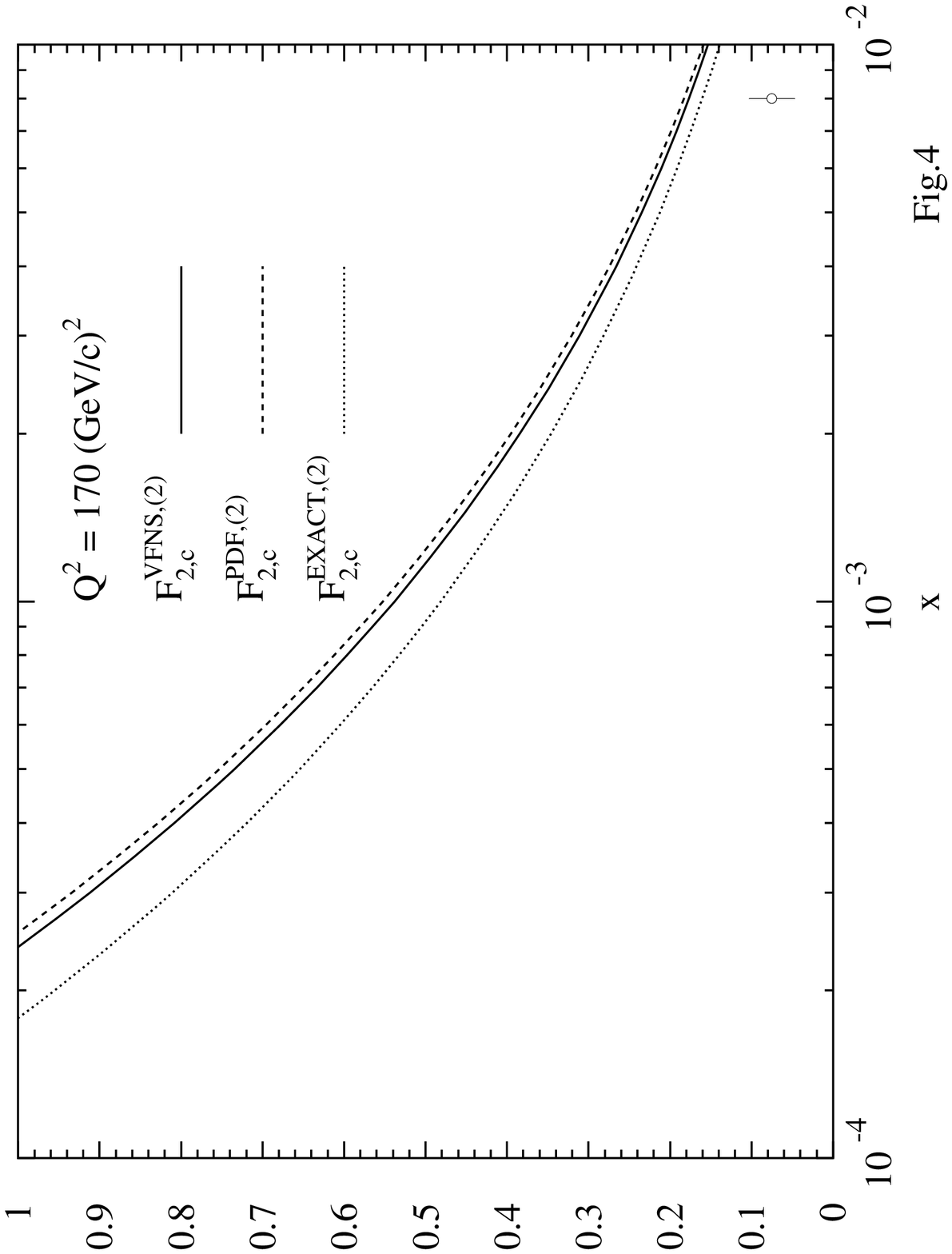}}
\end{picture}
\end{center}
\end{figure}
\end{document}